\documentclass[prd,aps,a4paper,nofootinbib,twocolumn]{revtex4} 

\newif\ifusesec
\usesectrue  
   
\usepackage{graphicx} 
\usepackage{mathrsfs}
\usepackage{amssymb}
\usepackage{latexsym,amsmath,amsfonts}
\usepackage{multirow}

\newcommand{\beq}{\begin{equation}}
\newcommand{\eeq}{\end{equation}}
\def\rightcontract{\mathop{\hbox{\vrule width0.5pt height6pt
  \vrule height0.5pt width6pt}}}
\begin{document}

\title{Deviation and precession effects in the field of a weak gravitational wave}

\author{Donato Bini}
\email{donato.bini@gmail.com}
\affiliation{Istituto per le Applicazioni del Calcolo \lq\lq M. Picone," CNR I-00185 Rome, Italy}
\affiliation{ICRANet, Piazza della Repubblica 10, I-65122 Pescara, Italy}

\author{Andrea Geralico}
\email{andrea.geralico@gmail.com}
\affiliation{Istituto per le Applicazioni del Calcolo \lq\lq M. Picone," CNR I-00185 Rome, Italy}
\affiliation{ICRANet, Piazza della Repubblica 10, I-65122 Pescara, Italy}

\author{Antonello Ortolan}
\email{ortolan@lnl.infn.it}
\affiliation{INFN - National Laboratories of Legnaro, Viale dell'Universit\`a 2, I-35020 Legnaro (PD), Italy}

\date{\today}

\begin{abstract}
Deviation and precession effects of a bunch of spinning particles in the field of a weak gravitational plane wave are studied according to the Mathisson-Papapetrou-Dixon (MPD) model.
Before the passage of the wave the particles are at rest with associated spin vector aligned along a given direction with constant magnitude.
The interaction with the gravitational wave causes the particles to keep moving on the 2-plane orthogonal to the direction of propagation of the wave, with the transverse spin vector undergoing oscillations around the initial orientation.
The transport equations for both the deviation vector an spin vector between two neighboring world lines of such a congruence are then solved by a suitable extension of the MPD model off the spinning particle's world line.
In order obtain measurable physical quantities a \lq\lq laboratory'' has been set up by constructing a Fermi coordinate system attached to a reference world line. The {\it exact} transformation between TT coordinates and Fermi coordinates is derived too.   
\end{abstract}

\maketitle

\section{Introduction}

In general relativity the relative acceleration between two neighboring world lines of test particles moving along geodesic orbits in a given background spacetime is described by the geodesic deviation equation. The rate of variation of the deviation vector connecting the two world lines is proportional to the Riemann tensor. 
If the particles are endowed with structure (spin, quadrupole moment, etc.), their world lines are no more geodesics, but accelerated due to the coupling between the spin and higher multipole moments with the curvature of the background.
This coupling will enter the associated generalized world line deviation equation as well.   

In this paper we consider a bunch of spinning test particles, which are described in terms of 4-momentum vector and spin tensor according to the Mathisson-Papapetrou-Dixon (MPD) model \cite{Mathisson:1937zz,Papapetrou:1951pa,Dixon:1970zza}.
The collection of particles forms a congruence of curves, each labeled by the position in space of any base point.
Imagine that one would like to compare momentum and spin of two different world lines of the congruence. This operation necessitates some transport law of quantities defined on a world line onto the other. Differently, if one is only interested in understanding the deviation of one world line with respect to the other, namely the relative position of the two, such transport laws are not needed and (only) the generalization of the geodesic deviation equation to accelerated families of curves is enough.
While the latter problem is well studied in the literature (especially in the case of a family of geodesics) the former is not so popular, even in familiar spacetimes of physical interest.

The aim of the present paper is to study such a problem in the case in which the background spacetime is that of a weak gravitational plane wave (GPW) and the congruence of accelerated world lines is that of spinning test particles.
The motion of particles with spin in a GPW spacetime has been investigated in Refs. \cite{Mohseni:2000re,Mohseni:2001sr}, later generalized to the case of small extended bodies also endowed with quadrupolar structure in Ref. \cite{Bini:2009zz}.
Studying relative motion between two of such particles requires an extension of the MPD model off the particle world line, because both 4-momentum and spin vector are defined only along it. 
Such an extension leading to a generalized world line deviation equation has been developed in Refs. \cite{Mohseni:2002zg,Mohseni:2004br}.
However, the transport equation for the spin vector is not discussed at all there, which is instead the main focus of the present analysis.

Here we consider a bunch of spinning particles which are at rest before the passage of the wave with associated spin vector aligned along a given direction with constant magnitude.
We first investigate the response of such a system to the interaction with the wave, determining the motion of the single particle of the congruence as well as the evolution of the spin vector components. 
We then solve the deviation equation between two neighboring world lines and the transport equation of the spin vector giving the precession effect between two nearby spacetime points.
Finally, in order to get a physically meaningful interpretation of the measurement of relative motion we construct a Fermi coordinate system about a reference spinning particle's world line, which acts as a \lq\lq laboratory reference system.''
We take advantage of the high symmetrical character of the background spacetime as well as of the weak field approximation to derive the exact transformation between TT coordinates and Fermi coordinates, generalizing previous works for geodesic world lines \cite{Fortini:1982ncb,Nesterov:1999ix,Rakhmanov:2014noa}. 
Indeed, this can be achieved in closed form only for a few spacetimes around very special world lines \cite{Chicone:2005vn,Klein:2009gz}, while most applications use a series expansion approximation from the very beginning \cite{Leaute:1983su,Bini:2005xt,Ishii:2005xq,Bini:2008xw,Bini:2014fba,Bini:2014qoa}. 

The physical problem underlying this study concerns eventually new effects related to the interaction between gravitational waves and spinning test particles which are poorly (or even not at all) investigated up to now, and may lead to measurable effects in the near future.
In fact, deviation and precession effects are the goal of a novel type of gravitational wave detectors, based on magnetized samples placed at a some given distance within a common superconducting shield \cite{gnome,romalis}.
We will show that measuring the different orientation of the spin vectors of two neighboring such samples due to the passage of the wave may allow to estimate the product between the amplitude and the frequency of the gravitational wave.

The paper is organized as follows. 
In Sec. II we introduce the MPD model for spinning particles and the generalized deviation equation for accelerated world lines in a curved background.
These concepts are then applied in Sec. III to the case of a weak gravitational wave spacetime, associated with a monochromatic gravitational wave traveling along a fixed spatial direction. We discuss the motion as well as the spin deviation effects in such a case, completing the section with some indication of possible measurements of the deviation effects. In Appendix A we shortly recall the geodesics of this metric and in Appendix B we give the new result concerning the exact map between spacetime and Fermi coordinates adapted to the world line of a spinning body. Notation and conventions follow Ref. \cite{MTW}. 
The metric signature is chosen as $-+++$; Greek indices run from 0 to 3, whereas Latin ones run from 1 to 3.

\section{Spinning particles and world line deviation}

Let us consider a family of spinning test particles in a given gravitational field (to be specified later).
The particles form a congruence of accelerated world lines, whose evolution follows the MPD equations of motion
\begin{eqnarray}
\label{pap_eqs}
\frac{D p^\alpha}{d\tau_U}&=&-\frac12 R^\alpha{}_{\beta\gamma\delta}U^\beta S^{\gamma\delta}
\equiv F^\alpha_{\rm(spin)}\,,\nonumber\\
\frac{DS^{\alpha\beta}}{d\tau_U } &=& p^{[\alpha} U^{\beta]}\,,\nonumber\\
S^{\alpha\beta}p_\beta &=& 0\,.
\end{eqnarray}
As standard, $U$ denotes the unit timelike vector tangent to the particle's world line ${\mathcal C}_U$, $p$ (a timelike vector, not unitary) is the generalized momentum and $S^{\mu\nu}$ is the spin tensor.
Both $p$ and $S$ have support on the world line ${\mathcal C}_U$ parametrized by the proper time $\tau_U$, i.e., with parametric equations $x^\alpha=x^\alpha(\tau_U)$ so that $U^\alpha = dx^\alpha/d\tau_U$.
\footnote{If the particles have additional structure besides spin, e.g., quadrupolar, octupolar structure, Eqs. \eqref{pap_eqs} are modified by corresponding additional terms. 
It is therefore quite natural to study these equations {\it only} at the linear order in spin. When working within this approximation we will explicitly specify it.}

For convenience, let us introduce the unit timelike vector $\bar u$ associated with $p$, i.e.,
\beq
p=m\bar u\,,\qquad  p\cdot p=-m^2\,,
\eeq
which is generally distinct from $U$ (but having support only along it) and related to it by a boost, i.e., 
\beq
\bar u=\gamma(\bar u, U)[U+\nu(\bar u, U)]\,.
\eeq
Here  $m$ is an \lq\lq effective mass" of the spinning particle, which is in general different from the \lq\lq bare mass" of a non-spinning particle and not a conserved quantity along the path, due to the evolution of the spin.
The relative velocity $\nu(\bar u, U)$, following \cite{Tod:1976ud}, is given by
\beq
\nu(\bar u, U)^\beta=\frac{P(U)^\beta{}_{\mu}S^{\alpha\mu}F_\alpha{}_{\rm(spin)}}{m^2+S^{\alpha\mu}F_\alpha{}_{\rm(spin)}U_\mu}\,,
\eeq
where $P(U)=g+U\otimes U$ projects orthogonally to $U$.

It is also useful to introduce the spin vector
\beq
S^\alpha = \frac12 \eta(\bar u)^{\alpha\beta\gamma}S_{\beta\gamma}, 
\eeq
where $\eta(\bar u)^{\alpha\beta\gamma}=\bar u_\sigma \eta^{\sigma\alpha\beta\gamma}$ is obtained from the 1-volume 4-form $\eta^{\sigma\alpha\beta\gamma}$.
The vector $S$ is orthogonal to $\bar u$ by definition and its magnitude 
\beq
s^2=S^\alpha S_\alpha 
=\frac12 S_{\alpha\beta}S^{\alpha\beta}
\eeq
is a constant of motion, as it can be easily checked by differentiating both sides along $U$ and using Eqs. \eqref{pap_eqs}.
Further constants of motion are associated with Killing vectors of the background spacetime. More precisely, if $\xi$ is a Killing vector, then 
\beq
J_\xi =\xi_\alpha p^\alpha+\frac12 S^{\alpha\beta}\nabla_{\beta} \xi_{\alpha}
\eeq
is a constant of the motion.

\subsection{World line deviation equation}

We are interested here in studying deviation effects among the particles of the congruence. 
Let $x^\alpha=x^\alpha (\tau_U,\sigma)$ denote the family of associated world lines, each curve labelled by $\sigma$ and parametrized by a proper time parameter $\tau_U$. As a consequence, the unit timelike tangent vector to the lines of the congruence is $U^\alpha_{\sigma}=\partial x^\alpha/\partial \tau_U$, whereas the deviation vector is $Y^\alpha_{(\tau_U)}=\partial x^\alpha/\partial \sigma$. 
The deviation vector $Y$ is  Lie-transported along $U$
\beq
\label{eq:lie}
\pounds_U Y=\nabla_U Y-\nabla_Y U=0\,.
\eeq
In addition,  $Y$ is not necessary orthogonal to $U$ and not unitary; therefore, in the Local Rest Space of $U$ (hereafter $LRS_U$) the \lq\lq effective" deviation vector or the position vector of nearby particles is its spatial projection
\beq
Y^\perp\equiv P(U)Y\,.
\eeq
From Eq. \eqref{eq:lie} it follows that
\beq
\nabla_U Y=\nabla_Y U\,, 
\eeq
so that differentiating (covariantly) again along $U$ this relation (as well as adding and subtracting terms properly), leads to
\begin{eqnarray}
\label{dev_eq}
\nabla_U \nabla_U Y&=&\nabla_U \nabla_Y U-\nabla_Y \nabla_U U+\nabla_Y \nabla_U U\nonumber\\
&=& [\nabla_U ,\nabla_Y ]U +\nabla_Y a(U)\nonumber\\
&=& R(U,Y)U +\nabla_Y a(U)\nonumber\\
&=& -{\mathcal K}(U)\rightcontract Y\,,
\end{eqnarray}
where we have used the Lie transport condition \eqref{eq:lie} in the Riemann tensor definition and we have introduced a \lq\lq strain tensor" \cite{Bini:2006vp}
\beq
{\mathcal K}_{\alpha\beta}(U)=-R_{\alpha\gamma \beta\delta}U^\gamma U^\delta -\nabla_\beta a(U)_\alpha\,, 
\eeq
built up with the tidal-electric Riemann tensor field, i.e., 
\beq
E(U)_{\alpha\beta}=R_{\alpha\gamma \beta\delta}U^\gamma U^\delta\,,
\eeq
and the covariant derivative of the acceleration spatial vector field $a(U)=\nabla_UU$ of the congruence $U$.
In the special case of $U$ geodesic we recognize the evolution equation for $Y$ as the geodesic deviation equation \eqref{dev_eq}, whose $3+1$ version was discussed in Refs. \cite{Bini:2006vp, Bini:2007zzb}.
The above considerations are general enough, in the sense that they hold for any family of world lines $U$.
Moreover, the geodesic deviation equation and its generalization to any accelerated family of world lines, Eq. \eqref{dev_eq}, allow to define the deviation vector $Y$ (and hence its spatial projection $P(U)Y$) {\it only} along $U$. In different words, taken a reference world line within the congruence $U$, the vector $Y$ is defined as an applied vector to any point along it. 
Note that Lie-transporting along $U$ an initially spatial vector with respect to $U$ does not lead in general to a transported vector which is still orthogonal to $U$. 

If one is interested in propagating the main tensors (momentum and spin) off the reference world line of the congruence $U$ the deviation vector plays an essential role. Indeed, one should consider the set of all spatial geodesics ${\mathcal C}_{\hat Y}$ with unit tangent vector $\hat Y=Y^\perp/||Y^\perp||$ emanating from a generic point $x^\alpha(\tau_{\rm (ref)})$ of the reference world line where $\hat Y=\hat Y(\tau_{\rm (ref)})$ (function of the proper time along the reference curve and defined only along it) and solve the parallel transport equations along these curves.
Their parametric equations $x^\alpha=y^\alpha(\sigma,\tau_{\rm (ref)})$ in terms of the affine arclength parameter $\sigma$ are the solutions of the following equations
\beq
\frac{d^2y^\alpha}{d\sigma^2}+\Gamma(y^\mu (\sigma))^\alpha{}_{\beta\gamma}\frac{dy^\beta}{d\sigma}\frac{dy^\beta}{d\sigma}=0\,,
\eeq
with initial conditions
\beq
y^\alpha\vert_{\sigma=0}=x^\alpha(\tau_{\rm (ref)})\,, \qquad
\frac{dy^\beta}{d\sigma}\bigg\vert_{\sigma=0}=\hat Y^\beta(\tau_{\rm (ref)})\,.
\eeq
Once the curve (the family of curves) ${\mathcal C}_{\hat Y}$ with tangent vector $\hat Y(\tau_{\rm (ref)},\sigma)\equiv{dy^\beta}/{d\sigma}$ is explicitly obtained, the solution of the parallel transport equations along it,
\beq
\label{partransY}
\nabla_{\hat Y}T^{\alpha\beta\ldots}=0\,,
\eeq
leads to the corresponding extended quantities off the reference world line, i.e., $T^{\alpha\beta\ldots}(\sigma,\tau_{\rm (ref)})$.
Subsequently,  for fixed values of $\sigma$, one can evolve these tensors along $U$, i.e., along any other curve of the congruence $U$, and compare with the corresponding quantities already existing there.

The Riemann tensor identities for the commuting vector fields $U$ and $Y$, i.e.,
\beq
[\nabla_U,\nabla_Y]C^\mu=R^\mu{}_{\nu \alpha\beta}C^\nu U^\alpha Y^\beta
%=R^\mu{}_{{\mathbf C} {\mathbf A}{\mathbf B}}
\,,
\eeq
imply
\begin{eqnarray}
0&=& \nabla_{Y} \nabla_U T^{\alpha\beta\ldots} +R^\alpha{}_{\mu\gamma\delta}T^{\mu\beta\ldots} U^\gamma {Y}^\delta \nonumber\\
&& +R^\beta{}_{\mu\gamma\delta}T^{\alpha\mu\ldots} U^\gamma {Y}^\delta +\ldots
\,,
\end{eqnarray}
which give in turn compatibility conditions for the corresponding extended fields.

In this paper, we will assume that the particles of the congruence $U$ are spinning test particles, while the general formulas derived above may involve any kind of acceleration.

\section{Deviation effects in the field of a weak gravitational wave}

Let us consider as a background spacetime that of a single gravitational wave travelling along the $x$ axis, with line element
\begin{eqnarray}
\label{GWmet}
ds^2&=& -dt^2+dx^2 +(1-h_+)dy^2 +(1+h_+)dz^2\nonumber\\
&& -2h_\times dy dz\,,
\end{eqnarray}
with $h_{+,\times}$ depending on $t-x$ only. 
In terms of the null coordinates
\beq
u= t-x \,,\quad
v= t+x \,,
\eeq
such that 
\beq
\partial_u =\frac{1}{2}(\partial_t -\partial_x)\,,\quad
\partial_v =\frac{1}{2}(\partial_t +\partial_x)\,,
\eeq
the metric becomes 
\begin{eqnarray}
ds^2&=& -du dv  +(1-h_+)dy^2 +(1+h_+)dz^2\nonumber\\
&& -2h_\times dy dz\nonumber\\
&=& -du dv +g_{AB}dx^A dx^B\,,
\end{eqnarray}
where $x^A$, $A=2,3$ denote the coordinates on the wave front.
The analysis of any kind of motion in this spacetime is simplified because of the existence of the three Killing vectors $\partial_x$, $\partial_y$, $\partial_v$.

Hereafter we will consider a monochromatic wave with
\beq
h_+(u)=A_+ \sin \omega u\,,\quad 
h_\times(u) =A_\times \cos \omega u \,,
\eeq
and limit our analysis to the linear order in $h_+$, $h_\times$.
Linear polarization corresponds to either $A_+= 0$ or $A_\times = 0$, whereas
circular polarization is assured by the condition $A_+ =\pm A_\times $. In general, one uses the \lq\lq polarization angle" $\psi= \tan^{-1}(A_\times/A_+)$.

Let us fix a family of fiducial observers at rest with respect to the $(t,x,y,z)$ coordinate grid, with four-velocity
\beq
n=\partial_t\,.
\eeq
An adapted triad to $n$ is given by
\begin{eqnarray}
\label{frame}
e_{\hat x}&=& \partial_x\nonumber\\
e_{\hat y}&=& \left(1+\frac12 h_+\right)\partial_y +\frac12 h_\times \partial_z \nonumber\\
e_{\hat z}&=&  \frac12 h_\times \partial_y+\left(1-\frac12 h_+\right)\partial_z \,.
\end{eqnarray}
This triad deals \lq\lq symmetrically" with  $e_{\hat y}$ and $e_{\hat z}$ and has a special geometrical meaning, since each leg of the triad undergoes a Fermi-Walker transport along $n$, namely
\beq
P(n)\nabla_n e_{\hat a}=0\,,
\eeq 
with $P(n)=g+n\otimes n$ projecting orthogonally to $n$. 
Moreover, the dependence on the coordinates of these frame vectors is only through $t-x$, implying a trivial Lie transport along any direction on the wave front.

It is customary to define the two polarization tensors
\begin{eqnarray}
e_+ &=& \partial_y\otimes \partial_y -  \partial_z\otimes\partial_z \nonumber\\
e_\times &=& \partial_y\otimes \partial_z + \partial_z\otimes \partial_y\,,
\end{eqnarray}
with associated notation
\beq
e_+\rightcontract X=X_+\,,\qquad
e_\times\rightcontract X=X_\times\,,
\eeq
where  the symbol $\rightcontract$ denotes right contraction between a tensor and a vector. In components, for example,
$[e_+\rightcontract X]_i=(X_+)_i=(e_+)_{ij}X^j$, that is
\beq
(X_+)_i=\delta_i^2 X_2-\delta_i^3 X_3\,,
\eeq
or
\begin{eqnarray}
X_+&=& X_2\partial_y -X_3\partial_z\nonumber\\
X_\times &=&X_3\partial_y +X_2\partial_z \,.
\end{eqnarray}

\subsection{Spinning particle motion}

The motion of an extended body in the spacetime of a weak gravitational wave \eqref{GWmet} has been studied in Refs. \cite{Mohseni:2000re,Mohseni:2001sr,Bini:2009zz}.
In the case of a spinning test particle and working at linear order in spin large simplifications arise.
For instance, in this case $p=m U$, with $m$ constant along $U$. 
We recall below the main results corresponding to the case of a spinning particle initially (before the passage of the wave) at rest at the origin of the coordinates,  with constant spin components.
The solution for the orbit $U^\alpha =dx^\alpha/d\tau_U$ is the following
\begin{eqnarray}
t(\tau_U)&=& \tau_U \,,\qquad
x(\tau_U)=0 \nonumber\\
y(\tau_U)&=&   \frac12 A_\times S_2^0[\cos(\omega\tau_U)-1]\nonumber\\
&&-\frac12  A_+ S_3^0[\sin(\omega\tau_U)-\omega \tau_U]   \nonumber\\
z(\tau_U)&=& -\frac12 A_\times S_3^0[\cos(\omega\tau_U)-1]\nonumber\\
&& -\frac12 A_+ S_2^0[\sin(\omega\tau_U)-\omega \tau_U]\,,
\end{eqnarray}
where $S_i^0$ ($i=1,2,3$) are the values of the coordinate components of the particle's spin at $\tau_U=0$.
In compact form
\begin{eqnarray}
x^A(\tau_U)&=&  \frac12 A_\times  [\cos(\omega\tau_U)-1] (S^0_+)^A\nonumber\\
&-& \frac12  A_+ [\sin(\omega\tau_U)-\omega \tau_U] (S^0_\times)^A\,.
\end{eqnarray}
Similarly
\begin{eqnarray}
\label{U_def}
U&=&\partial_t-\frac12\omega  A_\times S^0_+ \sin(\omega \tau_U)\nonumber\\
&&
-\frac12\omega  A_+ S^0_\times (\cos(\omega\tau_U)-1)
\,,
\end{eqnarray}
where we have chosen the location of the particle to be the origin at $\tau_U=0$.

The coordinate components of the spin vector are given by
\begin{eqnarray}
\label{Ssols}
S^t(\tau_U) &=& 0\,,\qquad
S^x(\tau_U) = S_1^0\,,\nonumber\\
S^y(\tau_U) &=& S_2^0+\frac12 S_2^0 A_+\sin(\omega\tau_U)\nonumber\\
&&+\frac12 A_\times S_3^0[\cos(\omega\tau_U)-1]\,,\nonumber\\
S^z(\tau_U) &=& S_3^0-\frac12 S_3^0 A_+\sin(\omega\tau_U)\nonumber\\
&& +\frac12 A_\times S_2^0[\cos(\omega\tau_U)-1]\,,
\end{eqnarray}
so that
\beq
S=S^0
+\frac12 A_+ S^0_+ \sin(\omega \tau_U)
+\frac12 A_\times S^0_\times (\cos(\omega\tau_U)-1)\,.
\eeq

\subsection{Deviation effects}

From these results, a straightforward computation gives the acceleration of $U$ (i.e., the spin force per unit mass) 
\begin{eqnarray}
a(U)&= &-\frac12 \omega^2\left[ \left(-S_3^0 A_+\sin(\omega\tau_U)+ S_2^0A_\times\cos(\omega\tau_U) \right)\partial_y \right.\nonumber\\
       && \left. -\left(S_2^0 A_+\sin(\omega\tau_U)+ S_3^0A_\times\cos(\omega\tau_U)\right) \partial_z\right] \nonumber\\
&=& -\frac12 \omega^2\left[ -A_+\sin(\omega\tau_U) S^0_\times 
+  A_\times\cos(\omega\tau_U)  S^0_+\right]\,.\nonumber\\
\end{eqnarray}
as well as the associated derivative
\begin{eqnarray}
\nabla_Y a(U) &=& \frac12 \omega^3 Y^{0\,u} \left[
\left(S_3^0 A_+\cos(\omega \tau_U)\right.\right.\nonumber\\
&&\left.\left.
+ S_2^0A_\times\sin(\omega\tau_U)\right)\partial_y \right.\nonumber\\
&&  \left.  
- \left(-S_2^0 A_+\cos(\omega\tau_U)+S_3^0 A_\times \sin(\omega\tau_U)\right) \partial_z \right]\nonumber\\
&=& \frac12 \omega^3 Y^{0\,u} \left[ A_+\cos(\omega \tau_U)S^0_\times
+  A_\times \sin(\omega\tau_U) S^0_+\right]\,,\nonumber\\
\end{eqnarray}
where $Y^{0\,u}=Y^{0\,t}-Y^{0\,x}$, and the only nonvanishing components of the (symmetric and tracefree) electric part of the Riemann tensor are
\begin{eqnarray}
E(U)_{xx}&=&-\frac12 \omega^2 A_+ \sin(\omega \tau_U)=-\frac12 \omega^2 h_+=-E(U)_{yy}\,,\nonumber\\
E(U)_{xy}&=& -\frac12 \omega^2 A_\times  \cos(\omega \tau_U)=-\frac12 \omega^2  h_\times \,.
\end{eqnarray}
In tensorial form
\beq
E(U)=-\frac12 \omega^2 \sum_{p=+,\times }  h_p e_p\,,
\eeq
where here and below $h_p$ is meant to be evaluated along the orbit.

Finally, the solution for the deviation vector is given by
\begin{eqnarray}
Y&=& Y^0\nonumber\\
&-&
\frac12\omega Y^{u\,0}\left[S^0_{\times}A_+(\cos(\omega \tau_U)-1) 
+S^0_{+}A_\times\sin(\omega \tau_U)\right]\,,\nonumber\\
\end{eqnarray}
where $Y^0=Y^{\alpha\,0} \partial_\alpha$ is the constant solution corresponding to the geodesic case and initial conditions have been chosen so that $Y(\tau_U=0)=Y^0$.
When expressed with respect to the frame \eqref{frame} it reads
\beq
Y=Y_{\rm (geo)}+Y_{\rm s}\,,
\eeq
with (see, e.g., Ref. \cite{MTW})
\begin{eqnarray}
Y_{\rm (geo)}&=& Y^{\alpha\,0}e_{\hat \alpha}\nonumber\\
&&
-\frac12\left[ A_\times Y^{z\,0} \cos(\omega \tau_U)
+A_+Y^{y\,0} \sin(\omega \tau_U)\right]e_{\hat y} \nonumber\\
&&
-\frac12\left[ A_\times Y^{y\,0} \cos(\omega \tau_U)
-A_+Y^{z\,0} \sin(\omega \tau_U)\right]e_{\hat z}\,,\nonumber\\
\end{eqnarray}
whereas the spinning part is given by
\begin{eqnarray}
Y_{\rm s}&=& -\frac12 \omega Y^{0\,u} \left[A_+ S^0_3  (\cos (\omega \tau_U)-1)
  +A_\times  S^0_2  \sin (\omega \tau_U) \right]e_{\hat y}\nonumber\\
&&
-\frac12 \omega Y^{0\,u} \left[A_+ S^0_2  (\cos (\omega \tau_U)-1)
  -A_\times  S^0_3  \sin (\omega \tau_U) \right]e_{\hat z}\,.\nonumber\\
\end{eqnarray}
Similarly
\beq
\hat Y=\hat Y_{\rm (geo)}+\hat Y_{\rm s}\,,
\eeq
where
\beq
\hat Y_{\rm (geo)}=\frac1{W}\left(1+\frac{\mathcal Z}{2W^2}\right)Y^{a\,0}\partial_a\,,
\eeq
with $W^2=\delta_{ab}Y^{a\,0}Y^{b\,0}$ and
\begin{eqnarray}
{\mathcal Z}&=&2Y^{y\,0}Y^{z\,0}A_\times \cos(\omega\tau_U)\nonumber\\
&&
+[(Y^{y\,0})^2-(Y^{z\,0})^2]A_+\sin(\omega\tau_U)\,,
\end{eqnarray}
and

\begin{widetext}

\begin{eqnarray}
\hat Y_{\rm s}&=&-\frac{\omega}{2W}({\mathcal T}\cdot Y^0)\left\{\partial_t+\frac{(Y^{x\,0})^2}{W^2}\left[
\partial_x-\frac{Y^{x\,0}}{(Y^{y\,0})^2+(Y^{z\,0})^2}\left(Y^{y\,0}\partial_y+Y^{z\,0}\partial_z\right)
\right]\right\}\nonumber\\
&&
-\frac{\omega}{2W}({\mathcal K}\cdot Y^0)\frac{Y^{x\,0}}{(Y^{y\,0})^2+(Y^{z\,0})^2}\left(Y^{z\,0}\partial_y-Y^{y\,0}\partial_z\right)
\,,
\end{eqnarray}
with the two vectors ${\mathcal T}$  and ${\mathcal K}$  given by
\begin{eqnarray}
{\mathcal T}&=&A_+ S^0_\times [\cos (\omega \tau_U)-1]+A_\times  S^0_+ \sin(\omega \tau_U)
\,,\nonumber\\
{\mathcal K}&=&A_+ S^0_+ [\cos (\omega \tau_U)-1]-A_\times S^0_\times \sin(\omega \tau_U) 
\,.
\end{eqnarray}

\end{widetext}

\begin{figure*}
\[
\begin{array}{cc}
\includegraphics[scale=0.3]{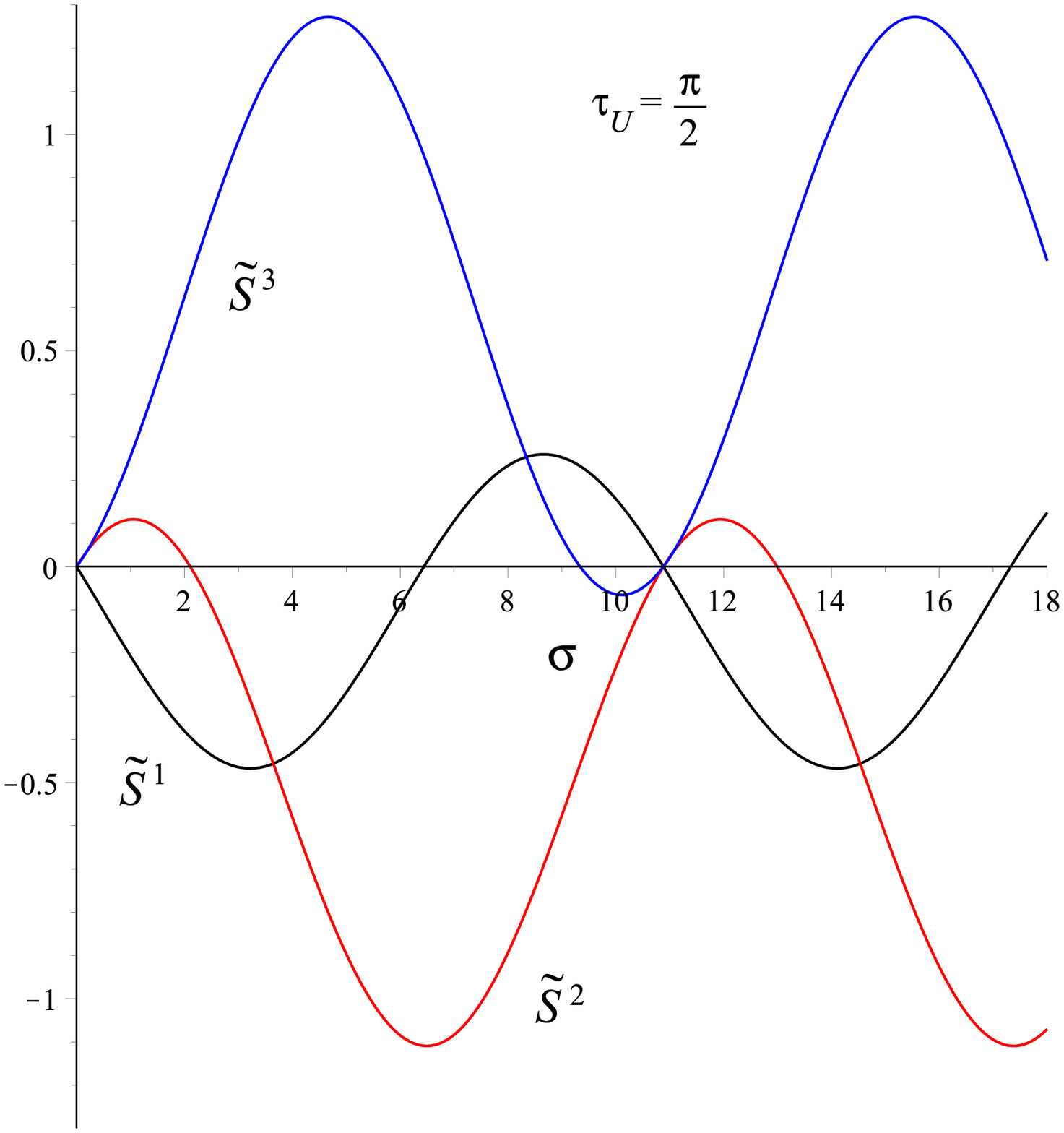}& 
\includegraphics[scale=0.3]{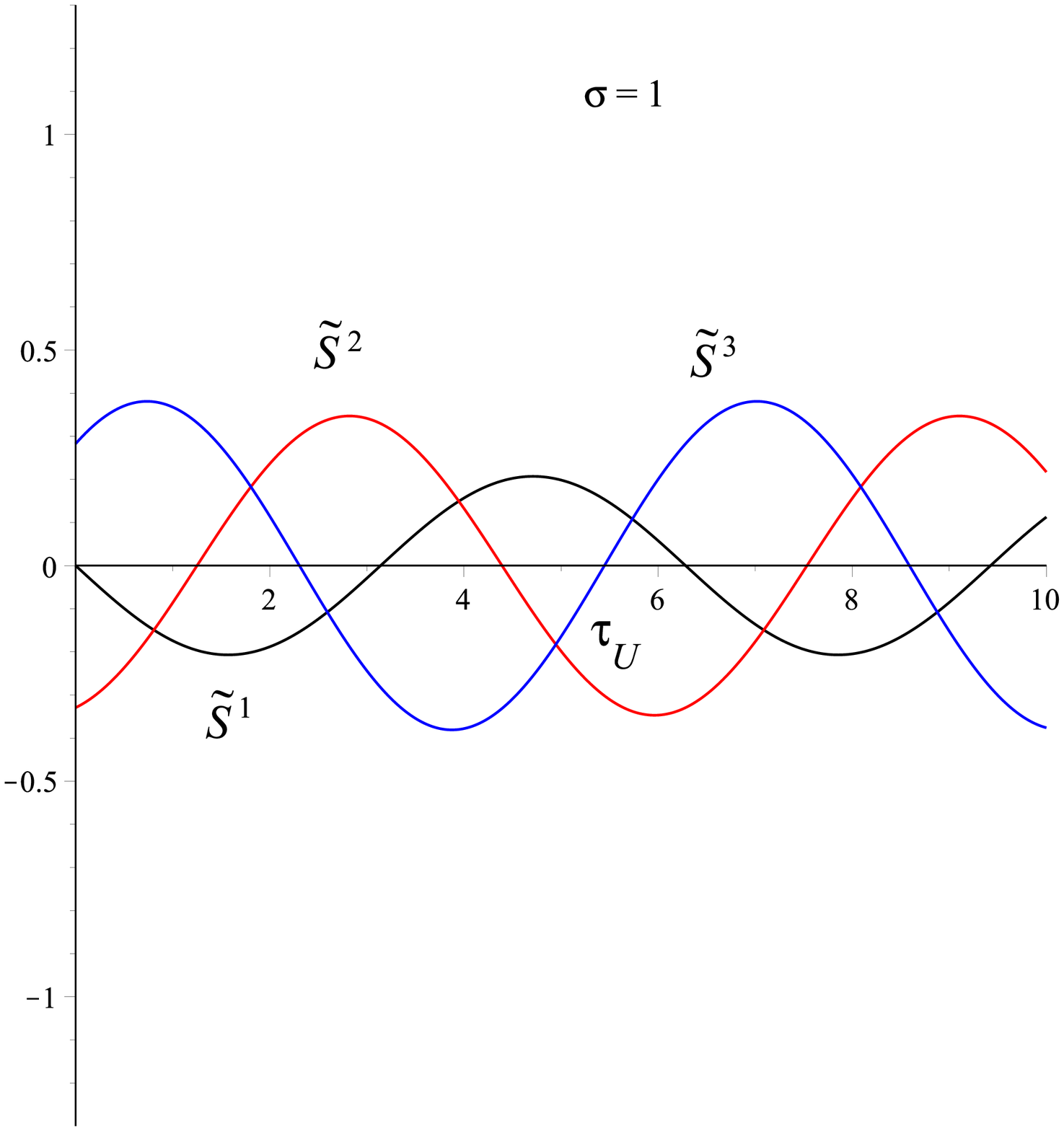}\cr
(a) & (b)
\end{array}
\]
\caption{The behavior of the spatial components $\tilde S^i(\tau_U,\sigma)$ of the transported spin vector as functions of $\sigma$ is shown in panel (a) for a fixed value $\tau_U=\pi/2$ of the proper time parameter along the world line of the spinning particle.
Panel (b) shows instead their behavior as functions of $\tau_U$ for a given displacement $\sigma=1$.
The choice of parameters is as follows: $S_1^0=s_0\sin\theta_s\cos\phi_s$, $S_2^0=s_0\sin\theta_s\sin\phi_s$, $S_3^0=s_0\cos\theta_s$, with $s_0=1$, $\theta_s=\pi/4=\phi_s$, $Y^{a\,0}=0.1$, $\omega=1$, and $A_\times=A_+\tan\psi$, with $A_+=1$, $\psi=\pi/6$, as an example.
}
\label{fig:1}
\end{figure*}

\begin{figure*}
\[
\begin{array}{cc}
\includegraphics[scale=0.3]{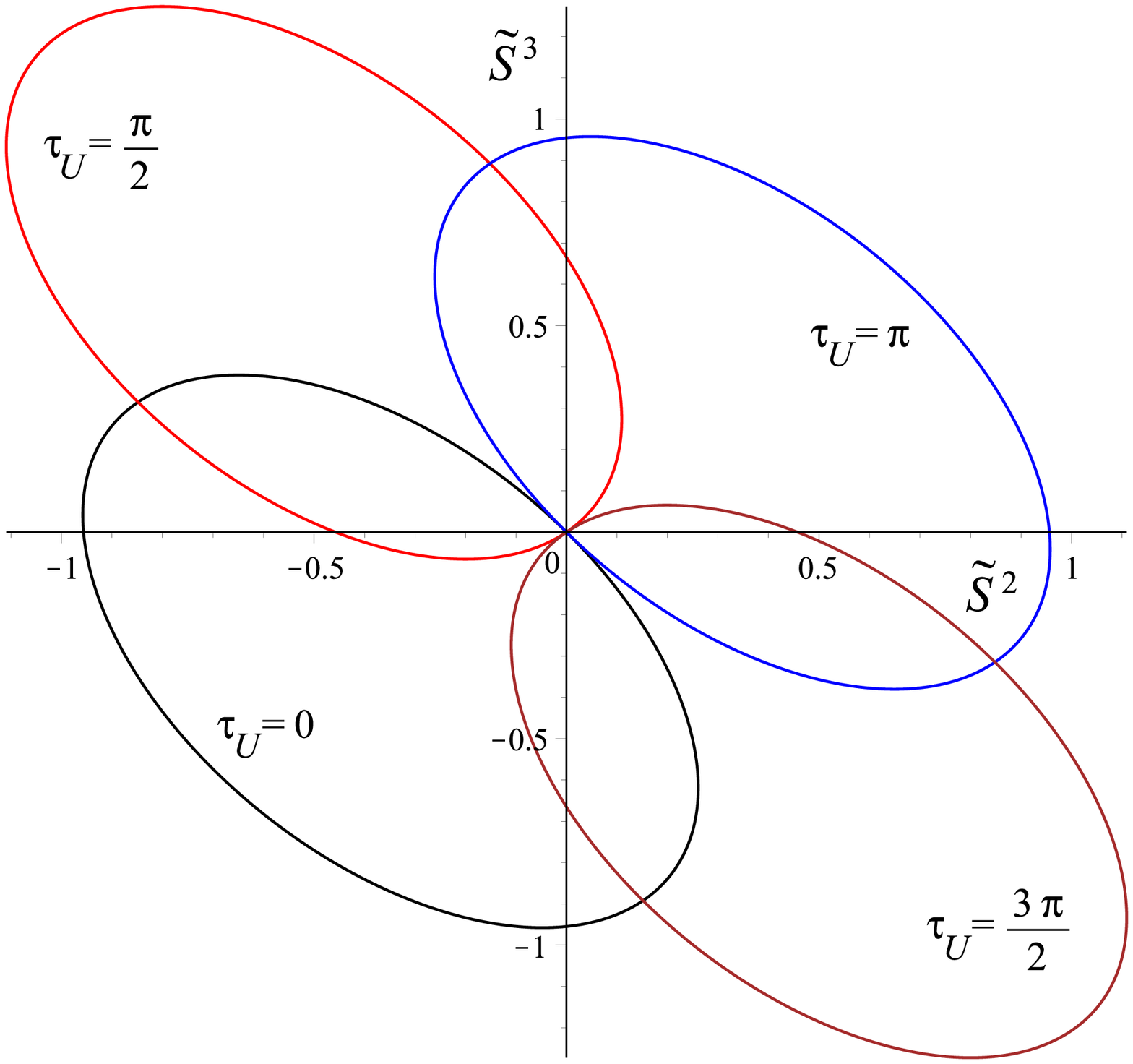}& 
\includegraphics[scale=0.3]{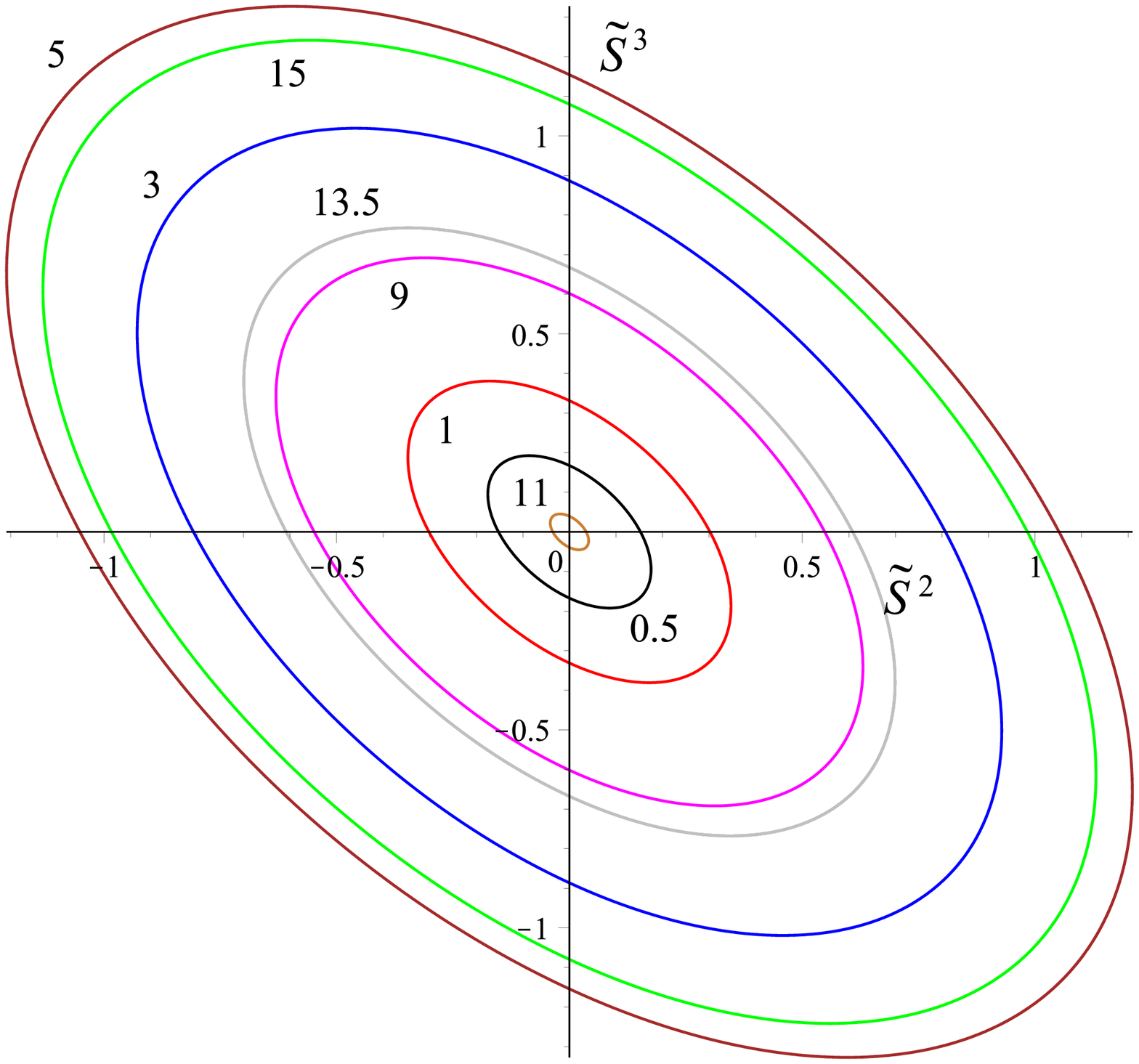}\cr
(a) & (b)
\end{array}
\]
\caption{The parametric curve $\tilde S^3(\tau_U,\sigma)$ versus $\tilde S^2(\tau_U,\sigma)$ is shown for different values of (a) $\tau_U$ and (b) $\sigma$ for the same parameter values as in Fig. \ref{fig:1}.
}
\label{fig:2}
\end{figure*}

Let us consider now the spatial geodesics emanating from a generic point on the world line $U$.
These are given in Appendix A (for $\mu^2=-1$ and $\lambda=\sigma$) and the values of the constants there correspond to 
\beq
\label{geoconst}
\alpha=\frac{Y^{0\,y}}{W}\,,\qquad
\beta=\frac{Y^{0\,z}}{W}\,,\qquad
E=-\frac{Y^{0\,x}}{W}\,.
\eeq
The transport equations \eqref{partransY} can then  be solved and the solution for the transported spin vector is given by $S^\alpha(\tau_U,\sigma)=S^\alpha(\tau_U)+\tilde S^\alpha(\tau_U,\sigma)$, with $S^\alpha(\tau_U)$ given by Eq. \eqref{Ssols} and
\begin{eqnarray}
\tilde S^1(\tau_U,\sigma)&=&(\beta S^0_2+\alpha S^0_3){\mathcal C}+(\alpha S^0_2-\beta S^0_3){\mathcal S}
\,,\nonumber\\
\tilde S^2(\tau_U,\sigma)&=&(-\beta S^0_1+E S^0_3){\mathcal C}+(-\alpha S^0_1+E S^0_2){\mathcal S}
\,,\nonumber\\
\tilde S^3(\tau_U,\sigma)&=&(-\alpha S^0_1+E S^0_2){\mathcal C}+(\beta S^0_1-E S^0_3){\mathcal S}
\,,
\end{eqnarray}
with $\tilde S^0(\tau_U,\sigma)=\tilde S^1(\tau_U,\sigma)$, where 
\begin{eqnarray}
{\mathcal C}&=&\frac{A_\times}{2E}\left[\cos(\omega(\sigma E+\tau_U))-\cos(\omega\tau_U)\right]\,,\nonumber\\
{\mathcal S}&=&\frac{A_+}{2E}\left[\sin(\omega(\sigma E+\tau_U))-\sin(\omega\tau_U)\right]\,,
\end{eqnarray}
with $\alpha$, $\beta$ and $E$ given by Eq. \eqref{geoconst}.
Their behavior as functions of $\sigma$ is shown in Fig. \ref{fig:1} (a) for a fixed value of the proper time parameter $\tau_U$.
Fig. \ref{fig:2} (b) shows, instead, their behavior as functions of $\tau_U$ for a given displacement $\sigma$.

%%% figure 1

%%% figure 2

\subsection{Measuring the spin deviation}

A physical measurement in a gravitational field by a given observer necessitates a continuous locally inertial system along the world line of the observer to get a correct interpretation. This can be realized by setting up a Fermi coordinate system in the neighborhood of the observer's world line.
The transformation between the background coordinates $(t,x,y,z)$ and Fermi coordinates $(T, X, Y, Z)$ is explicitly derived in Appendix B.
Here we simply use the results to rewrite the deviation components of the transported spin vector in a form which is suitable for a measurement process.
We find $\tilde S^T=\tilde S^X$ and

\begin{widetext}

\begin{eqnarray}
\label{Sfermic}
\tilde S^X
&=&-\frac1{2X}\left\{(S^0_2Y-S^0_3Z)A_+[\sin(\omega(T-X))-\sin(\omega T)]
+(S^0_2Z+S^0_3Y)A_\times[\cos(\omega(T-X))-\cos(\omega T)]\right\}
\,,\nonumber\\
\tilde S^Y&=&\frac1{2X}\left\{(S^0_1Y+S^0_2X)A_+[\sin(\omega(T-X))-\sin(\omega T)]
+(S^0_1Z+S^0_3X)A_\times[\cos(\omega(T-X))-\cos(\omega T)]\right\}
\,,\nonumber\\
\tilde S^Z&=&-\frac1{2X}\left\{(S^0_1Z+S^0_3X)A_+[\sin(\omega(T-X))-\sin(\omega T)]
-(S^0_1Y+S^0_2X)A_\times[\cos(\omega(T-X))-\cos(\omega T)]\right\}
\,,
\end{eqnarray}
which imply
\beq
\tilde S^X X+\tilde S^Y Y +\tilde S^Z Z=\frac{S^0_1}{2X}\{
2YZA_\times[\cos(\omega(T-X))-\cos(\omega T)]
+(Y^2-Z^2)A_+[\sin(\omega(T-X))-\sin(\omega T)]
\}\,.
\eeq

\end{widetext}

Let us prepare our device in such a way that the test body is spinning around the $Z$-axis before the passage of the wave with constant spin, i.e., $S^0_1=0=S^0_2$ and $S^0_3\equiv s_0$.
The interaction with the gravitational wave causes the spin vector to acquire nonvanishing spatial components all varying with time and displacements along the three spatial directions as from Eq. \eqref{Sfermic}, which simplifies as
\begin{eqnarray}
\tilde S^X
&=&\frac{s_0}{2X}\{ZA_+[\sin(\omega(T-X))-\sin(\omega T)]\nonumber\\
&&
-YA_\times[\cos(\omega(T-X))-\cos(\omega T)]\}
\,,\nonumber\\
\tilde S^Y&=&\frac{s_0}{2}A_\times[\cos(\omega(T-X))-\cos(\omega T)]
\,,\nonumber\\
\tilde S^Z&=&-\frac{s_0}{2}A_+[\sin(\omega(T-X))-\sin(\omega T)]
\,.
\end{eqnarray}
It is easy to check then the following (exact) properties satisfied by the  spin components
\begin{eqnarray}
\label{property_nice}
&& \tilde S^X X+\tilde S^Y Y +\tilde S^Z Z=0\,,\nonumber\\
&& \frac{(\tilde S^Y)^2}{ A_\times^2}+\frac{(\tilde S^Z)^2}{ A_+^2}=s_0^2\sin^2 \left( \frac{\omega X}{2}\right)\,,
\end{eqnarray}
in which the dependence on $T$ has disappeared. A nice geometrical interpretation of Eqs. \eqref{property_nice} can be given either in the spin space, i.e., at fixed $X,Y,Z$ or in the configuration space, i.e., at fixed $\tilde S^X,\tilde S^Y,\tilde S^Z$.
In fact,  in the spin space, the transverse components of the spin  \lq\lq belong" to an ellipse which intersects a plane. Using the component $\tilde S^X$ as a parameter, the explicit solution for 
$\tilde S^Y,\tilde S^Z$ corresponds actually to a circle, recalling the mechanical properties of the ellipsoid of inertia of a rigid body.
Vice versa,  in the configuration space $X,Y,Z$, with the spin components taken as fixed, the above equation implies a fixed value for the $X$ and straight line in the $Y$-$Z$ plane, leading to an unexpected simple geometrical  characterization of an otherwise complicated situation. 

Sufficiently close to the the reference spinning particle's world line it is enough to take the expansion of the above expressions up to the first order in the spatial Fermi coordinates, which gives 
\begin{eqnarray}
\label{TIP}
\tilde S^X
&=&-\frac12\omega s_0[YA_\times \sin(\omega T)+ZA_+\cos(\omega T)] + O(2)
\,,\nonumber\\
\tilde S^Y&=&\frac12\omega s_0XA_\times \sin(\omega T) + O(2)
\,,\nonumber\\
\tilde S^Z&=&\frac12\omega s_0XA_+\cos(\omega T) + O(2)
\,.
\end{eqnarray}
If we consider a second spinning particle located at coordinates $(X,0,0)$ with the same spin vector of the particle in the origin, the second and third equations in (\ref{TIP}) describe a torque-induced relative precession, as the directions of the two spins define a generalized cone in space.
Let us focus on the component of the spin vector on the wave front, with magnitude
\beq
\tilde S^\perp=\sqrt{(\tilde S^Y)^2+(\tilde S^Z)^2}\,. 
\eeq
It is worth noticing that for a circularly polarized  GPW $\tilde S^\perp$ is the amplitude of the relative precession cone induced by a GPW 
on the two spinning particles.
It turns out that to the first order in the distance (omitting the $O(2)$ notation too) 
\footnote{Clearly, extending the present calculation to higher orders in the distance is a trivial task, and does not require special consideration.}
and taking $A_+= \pm A_\times = h$
\begin{eqnarray}
\tilde S^\perp (X)&=&\frac12 s_0 h \omega X\,.
\end{eqnarray}
In Earth based laboratories separated by a distance $X=L$ we have
\beq
\tilde S^\perp (L)=\frac12 s_0h  \omega L\, , 
\eeq
where  $O(2)$ expansion requires $\omega L/c   \lesssim 1 $ (here the speed of light $c$ is made explicit).
Therefore, measuring $\tilde S^\perp (L)$ at different distances $L$  leads to a direct information about the gravitational wave concerning the product of amplitude and frequency.

The present result is interesting especially because it is under current consideration a novel experimental scheme enabling the investigation of spin couplings \cite{gnome}, based on synchronous measurements of optical magnetometer signals from several devices operating in magnetically shielded environments placed at distant locations \cite{romalis,Kimball:2016dnx}.
The primary interest is to probe couplings between spins and \lq\lq exotic'' fields beyond the Standard Model, e.g., axion-like fields, by analyzing the correlation between signals from multiple, geographically separated magnetometers.
However, we argue that the same scheme can be equally applied to the kind of interaction under investigation here, the gravitational wave signal inducing a torque on atomic spins.   
In this way, the apparatus should be able to detect the magnetic-like part of a gravitational wave by comparing two distant samples of elementary spins, differently (and complementary)  to the LIGO/VIRGO interferometers which succeeded in measuring the electric-like part of a gravitational wave using free falling masses. 
Our present study provides the underlying theoretical framework for the prediction of observed signals by this new kind of detectors.

For instance, for the gravitational wave event GW150914 recently observed by LIGO \cite{Abbott:2016blz}, with maximum amplitude $h\sim10^{-21}$ at a frequency $\omega=150$ Hz, we find
${\tilde S^\perp}/{s_0}\sim2.5\times10^{-21}$, if the detector consists of two magnetometers separated by a distance of $L=10^4$ Km \cite{gnome}.

\section{Concluding remarks}

The observation of gravitational wave signals GW150914 \cite{Abbott:2016blz} and GW151226 \cite{Abbott:2016nmj} by LIGO,  besides opening new horizons in gravitational wave astronomy and astrophysics, 
provides important  tests  of  the  general theory of relativity in the strong field regime.  However, up to now, the two LIGO interferometers have observed only the electric-like part of the waves, measuring the induced displacements of their free falling mirrors. The observation of the magnetic-like part of a gravitational wave, i.e., the measurement of the induced effects on mass current or spinning particles is still missing. We have shown that by using spinning test particles one can identify observable effects associated with the variation of the polarization. 

We have considered a bunch of spinning test particles initially at rest before the passage of the wave (plane, monochromatic, transverse in the present analysis, for simplicity), with associated spin vector aligned along a given direction with constant magnitude.
The interaction with the gravitational wave causes the particles to keep moving on the 2-plane orthogonal to the direction of propagation of the wave.
The transverse components of the spin vector undergo oscillations around their initial orientation, whereas the component parallel to the direction of
the wave propagation does not change. 
By solving the transport equations for both the deviation vector an spin vector between two neighboring world lines of such a congruence we have shown that the   relative precession of two spins, assumed to be along the transverse direction of a Fermi frame, which is our \lq\lq laboratory'' system and separated by a distance L along the direction of propagation of the GPW) is given by $\frac 12 h \omega L$. Even if the corresponding amplitudes of the GPW detected by LIGO  seem not very promising, it is true that the technology to measure spin precessions on a network of optical magneometers already exists \cite{gnome} and will be refined in the next years. 

As a final remark, we observe  that $ \tilde S^\perp (L)$ could be also measured by comparing the magnetizations due to electron spins in two ferromagnetic samples, separated by a distance $L$, and magnetized by  parallel external magnetic fields.

\appendix

\section{Geodesics}

The geodesic of the metric \eqref{GWmet} are given by \cite{Bini:2000xj,Sorge:2001sq}

\begin{widetext}

\begin{eqnarray}
U_{(\rm geo)}&=&\frac{1}{2E}[(\mu^2+f+E^2)\partial_t+(\mu^2+f-E^2)\partial_x]\nonumber\\
&&+[\alpha(1+h_+)+\beta h_\times]\partial_y+[\beta(1-h_+)+\alpha h_\times]\partial_z\ ,
\end{eqnarray}
where $\alpha$, $\beta$ and $E$ are conserved Killing quantities, $\mu^2=1,0,-1$ corresponding to timelike, null and spacelike geodesics respectively, and
\beq
f=\alpha^2(1+h_+)+\beta^2(1-h_+) +2\alpha\beta h_\times\ .
\eeq
The corresponding parametric equations of the geodesic orbits are then easily obtained
\begin{eqnarray}
\label{geogw}
t(\lambda)&=&t_0+E\lambda+x(\lambda)-x_0
\,, \nonumber\\
x(\lambda)&=&x_0+(\mu^2+\alpha^2+\beta^2-E^2)\frac{\lambda}{2E}\nonumber\\
&&
-\frac{1}{\omega E^2}\left\{\frac12(\alpha^2-\beta^2)A_+[\cos\omega(E\lambda+t_0-x_0)-\cos\omega(t_0-x_0)]
-\alpha\beta A_\times[\sin\omega(E\lambda+t_0-x_0)-\sin\omega(t_0-x_0)]\right\}
\,, \nonumber\\
y(\lambda)&=&y_0+\alpha\lambda
-\frac{1}{\omega E}\bigg\{\alpha A_+[\cos\omega(E\lambda+t_0-x_0)-\cos\omega(t_0-x_0)]-\beta A_\times[\sin\omega(E\lambda+t_0-x_0)-\sin\omega(t_0-x_0)]\bigg\}
\,, \nonumber\\
z(\lambda)&=&z_0+\beta\lambda
+\frac{1}{\omega E}\bigg\{\beta A_+[\cos\omega(E\lambda+t_0-x_0)-\cos\omega(t_0-x_0)]+\alpha A_\times[\sin\omega(E\lambda+t_0-x_0)-\sin\omega(t_0-x_0)]\bigg\}
\,, 
\end{eqnarray}
where $\lambda$ is an affine parameter and $x^\alpha_0=x^\alpha(\lambda=0)$.

\section{Fermi coordinates}

Let us construct a Fermi coordinate system $(T, X, Y, Z) = (T, X^1, X^2, X^3)$ in a neighborhood of the (accelerated) world line $U$ of the spinning particle.
They are defined by 
\beq
T=\tau_U\,,\qquad 
X^i=\sigma(\xi\cdot F_i)|_Q\,,
\eeq
where $\xi=\frac{dx^\mu(\sigma)}{d\sigma}\partial_\mu$ is the unit vector tangent to the unique spacelike geodesic segment of proper length $\sigma$ connecting a generic point $Q$ on the particle's world line with a generic spacetime point $P$ near $Q$ and satisfying the condition $(\xi\cdot U)|_Q=0$, and $\{F_i\}$ is a Fermi-Walker spatial triad given by
\beq
F_1 =  e_{\hat x}\,, \qquad
F_2 = \nu^{\hat y} \partial_t +e_{\hat y}\,, \qquad
F_3 = \nu^{\hat z} \partial_t +e_{\hat z}\,.
\eeq
The velocity components can be identified directly from Eq. \eqref{U_def}, by re-writing $U=\partial_t+\nu^{\hat y}e_{\hat y}+\nu^{\hat z}e_{\hat z}$. 

By using the solution \eqref{geogw} for the spatial geodesics, the spatial Fermi coordinates around the point $Q$ turn out to be given by
\begin{eqnarray}
\label{fermicoords}
X&=&
\frac{\sigma}{\sqrt{1-\alpha_{(0)}^2-\beta_{(0)}^2}}\left[1-\alpha_{(0)}\alpha_{(1)}-\beta_{(0)}\beta_{(1)}
-\frac12(\alpha_{(0)}^2-\beta_{(0)}^2)A_+\sin(\omega\tau_U)-\alpha_{(0)}\beta_{(0)}A_\times\cos(\omega\tau_U)\right]
\,,\nonumber\\
Y&=&\sigma\left[\alpha_{(0)}+\alpha_{(1)}+\frac12\alpha_{(0)}A_+\sin(\omega\tau_U)+\frac12\beta_{(0)}A_\times\cos(\omega\tau_U)
\right]
\,,\nonumber\\
Z&=&\sigma\left[\beta_{(0)}+\beta_{(1)}+\frac12\alpha_{(0)}A_\times\cos(\omega\tau_U)-\frac12\beta_{(0)}A_+\sin(\omega\tau_U)
\right]
\,,
\end{eqnarray}
to first order in the gravitational wave amplitudes, where we have used the orthogonality condition between $\xi$ and $U$ at $Q$ to express the constant of motion $E=E_{(0)}+E_{(1)}$ in terms of $\alpha=\alpha_{(0)}+\alpha_{(1)}$ and $\beta=\beta_{(0)}+\beta_{(1)}$.
The arclength parameter $\sigma$ as well as the constants $\alpha$ and $\beta$ must then be expressed in terms of the background coordinates $(t,x,y,z)$.

We are interested here in the inverse transformation.
Therefore, solving Eq. \eqref{fermicoords} for $\sigma$, $\alpha$ and $\beta$ yields
\begin{eqnarray} 
\sigma&=&\sqrt{X^2+Y^2+Z^2}
\,,\nonumber\\
\sigma\alpha&=&Y-\frac12Z A_\times\cos(\omega T)-\frac12Y A_+\sin(\omega T)
\,,\nonumber\\
\sigma\beta&=&Z-\frac12Y A_\times\cos(\omega T)+\frac12Z A_+\sin(\omega T)
\,,
\end{eqnarray}
with in addition
\beq
\sigma E=-X-\frac12\omega A_+(S_2^0 Z+S_3^0 Y)[\cos(\omega T)-1]+\frac12\omega A_\times(S_2^0 Y-S_3^0 Z)\sin(\omega T)\,,
\eeq
which, once inserted in Eq. \eqref{geogw} with $x^\alpha_0=x^\alpha(\tau_U)$, finally gives the {\it exact} coordinate transformations
\begin{eqnarray}
t&=&T
-\frac12A_+\left\{\frac{Y^2-Z^2}{X}\left[\frac{\cos(\omega(T-X))-\cos(\omega T)}{\omega X}-\sin(\omega T)\right]+\omega(S_2^0 Z+S_3^0 Y)[\cos(\omega T)-1]\right\}\nonumber\\
&&
+A_\times\left\{\frac{YZ}{X}\left[\frac{\sin(\omega(T-X))-\sin(\omega T)}{\omega X}+\cos(\omega T)\right]-\frac12\omega(S_2^0 Y-S_3^0 Z)\sin(\omega T)\right\}
\,,\nonumber\\
x&=&X
-\frac12A_+\frac{Y^2-Z^2}{X}\left[\frac{\cos(\omega(T-X))-\cos(\omega T)}{\omega X}-\sin(\omega T)\right]\nonumber\\
&&
+A_\times\frac{YZ}{X}\left[\frac{\sin(\omega(T-X))-\sin(\omega T)}{\omega X}+\cos(\omega T)\right]
\,,\nonumber\\
y&=&Y
+A_+\left\{Y\left[\frac{\cos(\omega(T-X))-\cos(\omega T)}{\omega X}-\frac12\sin(\omega T)\right]-\frac12S_3^0[\sin(\omega T)-\omega T]
\right\}\nonumber\\
&&
-A_\times\left\{Z\left[\frac{\sin(\omega(T-X))-\sin(\omega T)}{\omega X}+\frac12\cos(\omega T)\right]-\frac12S_2^0[\cos(\omega T)-1]
\right\}
\,,\nonumber\\
z&=&Z
-A_+\left\{Z\left[\frac{\cos(\omega(T-X))-\cos(\omega T)}{\omega X}-\frac12\sin(\omega T)\right]+\frac12S_2^0[\sin(\omega T)-\omega T]
\right\}\nonumber\\
&&
-A_\times\left\{Y\left[\frac{\sin(\omega(T-X))-\sin(\omega T)}{\omega X}+\frac12\cos(\omega T)\right]+\frac12S_3^0[\cos(\omega T)-1]
\right\}
\,.
\end{eqnarray}
This transformation contains the new (spinning) terms additively, in the sense that it is formally of the type
\beq
x^\alpha =X^\alpha + f_{\rm (geo)}^\alpha (X^\mu)+ f_{\rm (spin)}^\alpha (X^\mu)\,.
\eeq
The spacetime metric in Fermi coordinates is then simply evaluated by
\beq
g_{AB}(X)=\frac{\partial x^\alpha}{\partial X^A}\frac{\partial x^\beta}{\partial X^B}g_{\alpha\beta}(x(X))\,.
\eeq
Up to the second order in the spatial Fermi coordinates its nonvanishing components are
\begin{eqnarray}
g_{TT}&=&-1-\left[S_2^0 Z+S_3^0 Y-\frac12(Y^2-Z^2)\right]\omega^2A_+\sin(\omega T)
+\left[S_2^0 Y-S_3^0 Z+YZ\right]\omega^2A_\times\cos(\omega T)
+O(3)
\,,\nonumber\\
g_{TX}&=&-\frac13\omega^2\left[(Y^2-Z^2)A_+\sin(\omega T)+2YZA_\times\cos(\omega T)\right]
+O(3)
\,,\nonumber\\
g_{TY}&=&\frac13\omega^2X\left[YA_+\sin(\omega T)+ZA_\times\cos(\omega T)\right]
+O(3)
\,,\nonumber\\
g_{TZ}&=&\frac13\omega^2X\left[-ZA_+\sin(\omega T)+YA_\times\cos(\omega T)\right]
+O(3)
\,,\nonumber\\
g_{XX}&=&1+\frac16\omega^2\left[(Y^2-Z^2)A_+\sin(\omega T)+2YZA_\times\cos(\omega T)\right]
+O(3)
\,,\nonumber\\
g_{XY}&=&-\frac12g_{TY}+O(3)\,,\qquad
g_{XZ}=-\frac12g_{TZ}+O(3)
\,,\nonumber\\
g_{YY}&=&1+\frac16\omega^2X^2A_+\sin(\omega T)
+O(3)
\,,\qquad
g_{YZ}=\frac16\omega^2X^2A_\times\cos(\omega T)
+O(3)
\,,\nonumber\\
g_{ZZ}&=&1-\frac16\omega^2X^2A_+\sin(\omega T)
+O(3)
\,,
\end{eqnarray}
in agreement with the series-expanded metric components in Fermi coordinates. 
Of course here we also have the exact (non-expanded) form of the metric, having identified the exact coordinate transformation from one coordinate system to the other.

\end{widetext}

\section*{Acknowledgments}
D.B. thanks ICRANet for partial support. A.O. thanks   Dr.  
M. L. Ruggiero for useful discussion.

\end{document}